\documentclass[a4paper, amsfonts, amssymb, amsmath, reprint, showkeys, nofootinbib,superscriptaddress,twocolumn ]{revtex4-2}
\usepackage{float}
\usepackage{graphicx}
\usepackage{longtable}
\usepackage[left=15mm,right=13mm,top=23mm,columnsep=15pt]{geometry} 
\usepackage[caption=false]{subfig}
\usepackage{hyperref} 
\usepackage{multirow}
\usepackage{array}
\usepackage{appendix}

\newcolumntype{L}{>{\centering\arraybackslash}m{6cm}}

\begin{document}
\title{Study of freeze-out dynamics of strange hadrons }
\author{Sushant K. Singh}
    \affiliation{Variable Energy Cyclotron Centre, 1/AF, Bidhan Nagar , Kolkata-700064, India}
    \affiliation{HBNI, Training School Complex, Anushakti Nagar, Mumbai 400085, India}
\author{Purabi Ghosh}
    \affiliation{Dept. of Applied Physics and Ballistics, F. M. University, Balasore-756019, India.}
    
\author{Jajati K. Nayak}
    \email[Correspondence email address: ]{jajati-quark@vecc.gov.in}
    \affiliation{Variable Energy Cyclotron Centre, 1/AF, Bidhan Nagar , Kolkata-700064, India}


\date{\today} 

\begin{abstract}
 We study the chemical freeze-out dynamics of strange particles ($K,\, \Lambda,\, \Sigma$) from a homogeneous and isotropically expanding hadronic system of $\pi, K, \rho, N, \Lambda$ and $\Sigma$ with zero net baryon density. We use the momentum integrated Boltzmann equation and study their evolution over the bulk hadronic matter, a condition being similar to the one created at top RHIC and LHC energies. The cross-sections, which are input to the equations, are taken either from phenomenological models or parameterized by comparing against experimental data. From this microscopic calculation we find that these strange particles freeze-out near transition temperature $\approx T_c$ due to large relaxation time. The continuous cease of the inelastic processes due to gradual fall in the temperature and decrease in the number density, thus lead to early freeze out of strange hadrons $K, \Lambda$ and $\Sigma$ which happens sequentially near $T_c$. However, freeze-out of these strange species near Tc appears as a sudden and simultaneous process, which is mostly predicted by thermal model while explaining the yield of identified particles at RHIC and LHC energies.  
\end{abstract}

\maketitle 

\section{Introduction}
Experimental observation and theoretical analysis suggest that quark gluon plasma(QGP)-one of the colored phases of quantum chromodynamics(QCD) is formed in relativistic heavy ion collisions at top RHIC and LHC energies~\cite{RHICwhitepaper05,kapusta03,gyulassy04,hotqcdwhite15,rafelski16,banerjee10,rafelski20,raju08}. The temperature and density of QGP is much higher compared to the normal nuclear matter. Once QGP is produced it undergoes a transition to hadronic matter-a color less QCD phase, when the temperature of the system cools down to $T_c$ due to expansion. 
The system of interacting hadrons undergoes further expansion which leads to the decoupling of various hadronic species along with the
decrease of temperature. In such expanding system the decoupling is decided by the scattering rate of interacting species and expansion rate of the system. At the decoupling or freeze out, the particles stop interacting chemically or kinetically.

It has been a long standing issue to analyse with rigor whether all hadron species decouple at the same time/temperature in an expanding system or they do gradually at different times/temperatures. It is intuitive that different particle species decouple from the medium at different temperatures as their masses and interaction crosssections are different. But sometimes the system dynamics is so interesting that the scattering rates of particles and expansion rate of the system compel us to infer that particles follow a common or simultaneous freeze out. Hence, it is of worth to study the freeze out behaviour of various species in different systems; more specifically in systems
produced at RHIC and LHC energies, as this study would help in summarizing the properties of QCD matter and mapping some portion of QCD phase diagram. 

The scenario where different particle species decouple gradually at different temperatures or times is refered as
sequential freeze-out. On the other hand, when they decouple at same time or temperature, the scenario is termed as simultaneous or common freeze out. Both sequential and common freeze outs are realised in relativistic heavy ion collisions in both cases of chemical and kinetic decouplings. 

In case of \emph{chemical freeze-out}, the inelastic scatterings between different hadron species stop, where as, in case of \emph{kinetic freeze-out}, the elastic scatterings cease following a  
free stream motion towards the detector. Exact informations of both chemical and kinetic freeze outs are important as hydrodynamic calculations, which are successful in describing the hot and dense fluid produced at RHIC and LHC, need these inputs. Most of the calculations\cite{hirano15,rupa20,rupa06,qian19,qianprc20,jknplb14,denicol18,gale13} of different observable such as net yield of identified hadron species, electromagnetic spectra, flow  of the different species {\it etc.} which provide important thermodynamic information of the produced system, assume simultaneous and sudden freeze-out scenario of these hadron species. Cooper-Frye formula is employed with assumption of ``sudden freeze-out`` which considers the mean free path of the hadron species becomes infinite suddenly through a thin freeze-out hyper surface.
\begin{figure*}
\centering
\subfloat{
   \includegraphics[height=6.0cm,width=7.5cm]{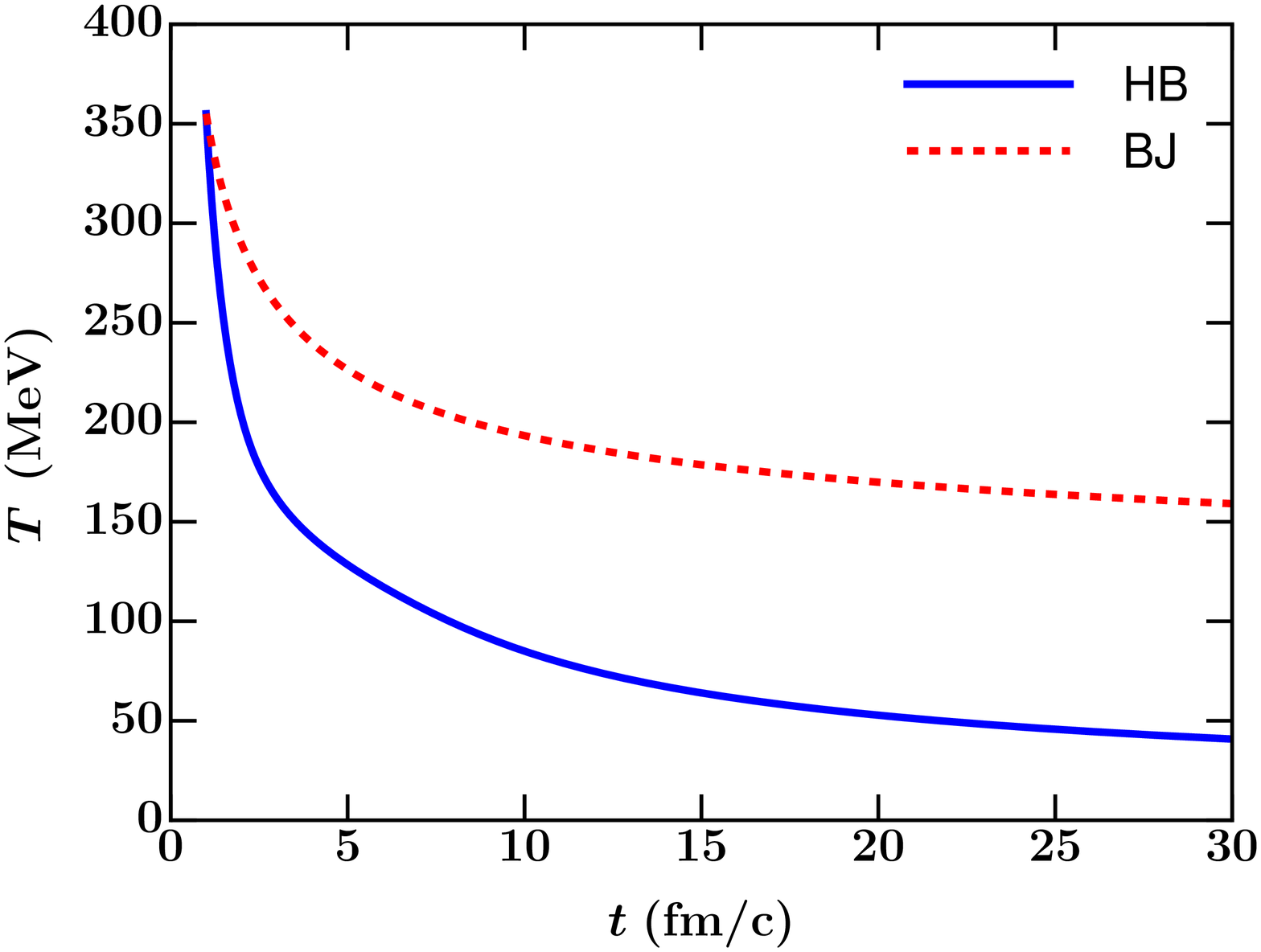}
 }
\subfloat{
   \includegraphics[height=6.0cm,width=8.5cm]{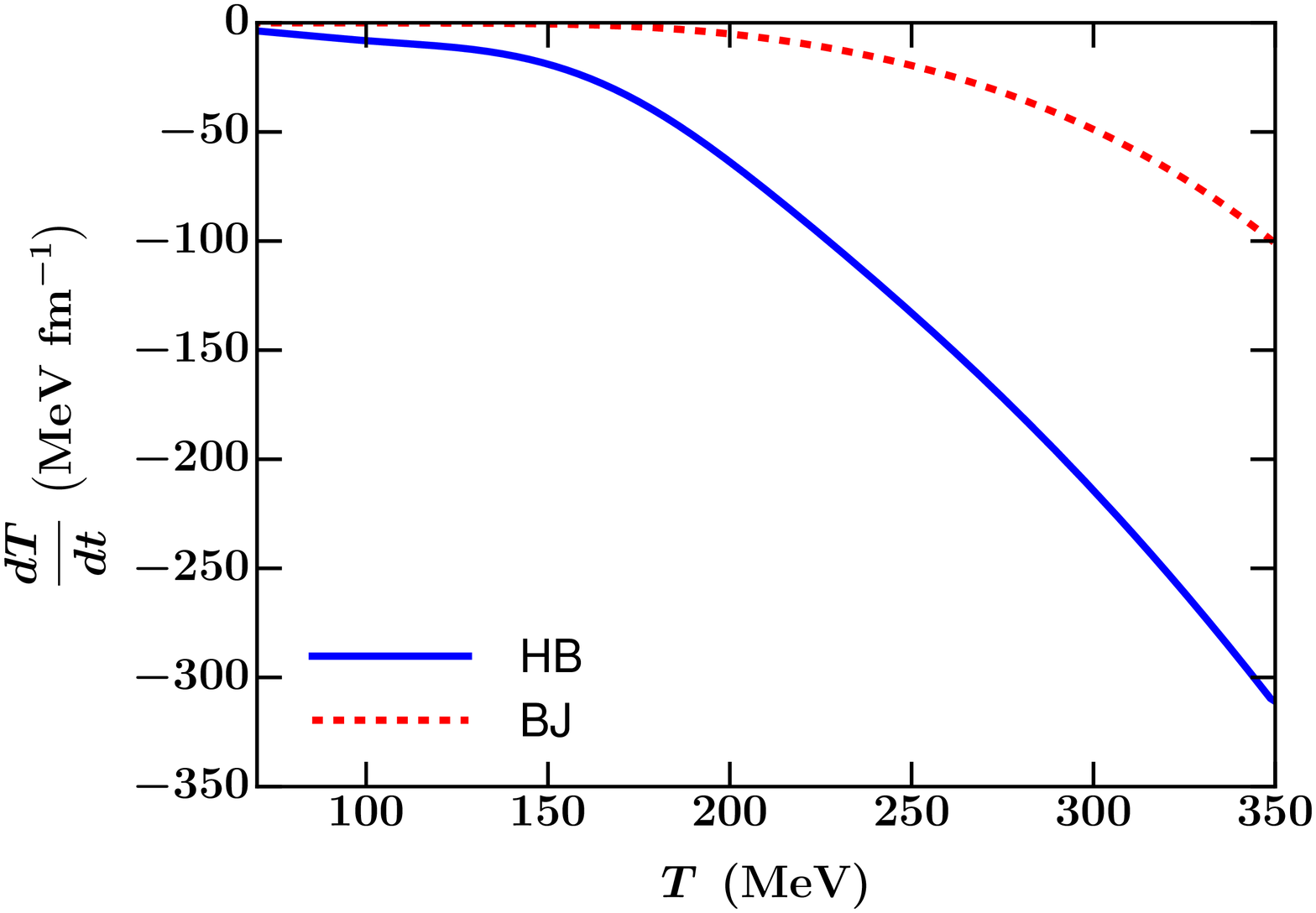}
 } 
\caption{(left panel) Temperature as a function of time and (right panel) $dT/dt$ as a function of temperature, for Bjorken (BJ) flow and Hubble-like (HB) flow. The initial temperature is chosen as $T_0 = 355\text{ MeV}$ at time $t_0 = 1\text{ fm/c}$. }
\label{fig_expansion_model}
\end{figure*}

Most importantly, the chemical freeze-out temperature is used to constrain the boundary conditions for the hydro dynamical evolution. For example, the electromagnetic spectra, which gives the information of initial temperature($T_i$) or energy density($\epsilon_i$) of hot dense QGP system, are evaluated using hydrodynamics and/or transport model which needs the accurate information of the kinetic as well as chemical freeze-out temperatures. In many hybrid-model calculations, transport approach is employed at transition temperature $T_c$ assuming a common chemical freeze-out for all species. Considering these pertinent issues, a thorough investigation is required to study freeze out dynamics which can provide temperature domain of hydrodynamic description accurately.

Hence it is highly important to investigate the freeze-out dynamics of the system produced in heavy ion collisions. In this article, we only focus on chemical freeze-out and discuss it for the strange hadrons $K, \Lambda, \Sigma$. In fact the chemical freeze-out dynamics is complex in nature due to the lack of complete understanding of all interactions in a multi-component hadronic fluid created at RHIC and LHC. It is really challenging as many inelastic scattering channels with different threshold energies are to be handled for studying one species. With the fall in temperature and dilution in density, inelastic channels stop interacting gradually. Which in fact reminds the continuous nature of chemical freeze-out processes.  

In this work, an attempt has been made to analyse microscopically the chemical freeze-out scenario of strange hadrons $K, \Lambda$ and $\Sigma$ using momentum integrated Boltzmann equation by considering number changing inelastic processes. We follow the approach as discussed in \cite{kolb_book} to study various species in case of early universe. The scattering rates have been calculated here and compared with the expansion rates following Bjorken and Hubble like dynamics. Then the decoupling of $K, \Lambda$ and $\Sigma$ have been analysed from both these rates. The present calculation provides a clue whether the assumption of sudden freeze out at top RHIC and LHC energies is valid or not. This approach microscopically considers the interaction of various channels and species and different from other attempts made earlier to study freeze out scenario ~\cite{kodama95plb,molnar2007,heinz2003,teany2002,heinz2006,kodama2004prl,urqmd18,sandeep17,bellwied18}. The author in ~\cite{molnar2007} studied the successive kinetic freeze-out in a Bjorken type of expansion and in ~\cite{heinz2003} the authors describe the breakdown of hydrodynamics and freeze-out of particles through Cooper-Frye prescription with the assumption of sudden transition of particles in the fluid element of perfect local thermal equilibrium to free streaming once criterion for kinetic freeze-out is achieved. Author in \cite{sandeep17} discuss about double freeze out scenario (chemical) using Hadron Resonance Gas model. Sequential freeze out has also been advocated in \cite{bellwied18} using lattice calculation.

In this article, we are dealing only with chemical freeze out. The study of kinetic freeze out, which happens later, is more complex to study using transport equation. This is because, the evolution of momentum distribution is far more complex than the evolution of particle densities. At this moment it may be worth to point out that the evolution of particular species after complete freeze out is simple: particle number density goes as $R^{-3}$ and momenta falls as $R^{-1}$ keeping the total number fixed, where $R$ is the system dimension. Since we are dealing with the study of chemical freeze-out, the equilibrium would mean as chemical equilibrium from the next. We would mention thermodynamic or kinetic equilibrium explicitly if phenomena related to both appears anywhere. Similarly the scattering here is basically inelastic scattering.  

In the next section we give a note on the chemical equilibrium and freeze-out scenario to set up the problem. In Sec. \ref{sec_pik} we discuss the dynamics by setting up rate equation with notations for two component system $\pi$, $K$  and discuss freeze out of $K$. We follow Ref~\cite{kolb_book} for the analysis. In section~\ref{sec_all}, we then study the chemical freeze-out of $K,\, \Lambda,\, \Sigma$. We finally summarize in section~\ref{sec_conclude}.
\section{Chemical equilibrium and freeze-out}
A system in complete thermodynamic equilibrium(chemical,mechanical and kinetic) may remain in equilibrium forever as long as the system is static and isolated. If any system consists of multiple species with some in complete equilibrium and some are not, then the species which are not in equilibrium would try to achieve it after certain time depending on their scattering rates. 

When a multi-component hot system is not in chemical equilibrium, then the particles do collide with other particles of same species and also with those of other species changing their number till the forward and backward reaction rates are same and the system achieves chemical equilibrium. The numbers in the species get fixed. However the particles may continue to collide with others, but elastically, without changing the number. The entire dynamics may be described by Boltzmann transport equation if the system is dilute enough to ignore three and higher body interactions.

In the above case of chemically un-equilibrated hot-dense static matter, system relaxes due to inelastic scatterings of various channels. But the relaxation processes become slower in case of expanding system as inter particle separation increases, thus lowering the scattering rate.
The system can maintain equilibrium till the scattering rate is equal or more than the expansion rate, otherwise not. If the equilibrium for a particular species of the multi-component fluid is concerned, then scattering rate of that species should be more compared to the rate of expansion. 

Following two scenarios may be possible in case of expanding system. Expansion with constant temperature (external pumping of the particles or temperature) or expansion with the variation of temperature. The decrease in temperature with expansion is a largely observed phenomenon for an isolated system. Such scenario is similar to the evolution of hot-dense matter produced in relativistic heavy ion collisions at RHIC and LHC  or the scenario encountered in the evolution of early universe. Gradually with increase in time and decrease in temperature, the scattering rate becomes comparable or less to the expansion rate and some of the inelastic channels may cease which marks the beginning of the phenomenon of chemical freeze-out. When all the processes involving the particles of one species cease then the freeze out of that species happens and corresponding temperature is called the chemical freeze out temperature($T_{ch}$ for that species. If the temperature and number density in the system is not sufficient initially then chemically unequilibrated system may start freezing out without going to equilibrium during expansion.
\begin{figure}[H]
\centering
\includegraphics[height=6.5cm,width=8.0cm]{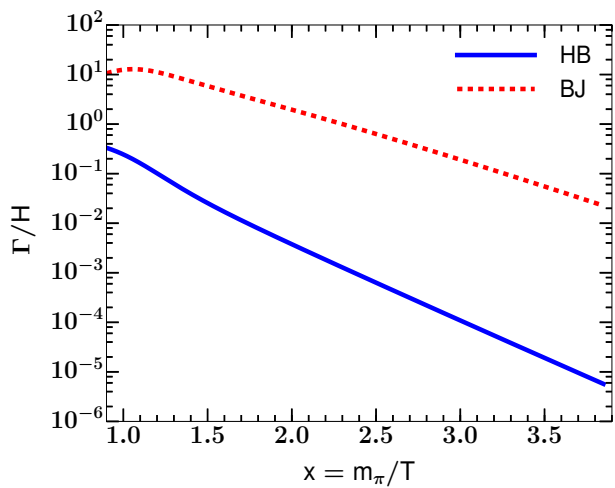}  
\caption{$\Gamma/H$ as a function of $x$ for Bjorken (BJ) flow and Hubble-like (HB) flow in $\pi-K$ system.}
\label{fig_ratio_pik}
\end{figure}

Let's call the temperature at which chemical freeze-out of a particular species starts as $T_{\text{chi}}$ {\it i.e.,} the cease of first inelastic process with involving the particles of that species. As time progresses and temperature falls, more and more inelastic channels stop and mean free path increases gradually. At certain temperature, $T_{\text{ch}}$, all the inelastic channels stop as the system can't provide the required threshold, the mean free path(considering inelastic cross sections) becomes infinite. That is the $T_{ch}$ of that species. Hence the entire process of chemical freeze-out happens in a temperature band $T_{\text{chi}}-T_{\text{ch}}$ continuously, not suddenly at any particular temperature.
\begin{figure}[H]
\centering
\includegraphics[height=6.0cm,width=7.5cm]{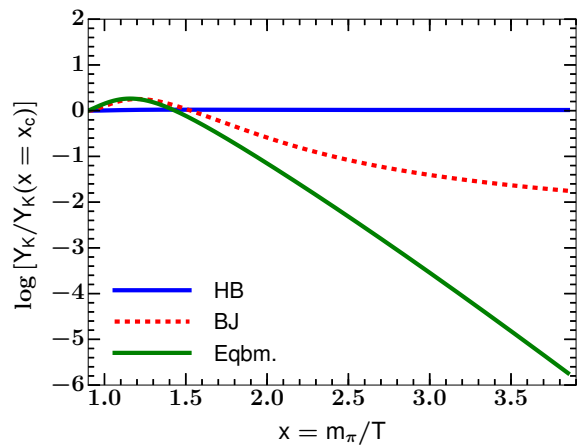}  
\caption{Numerical solution of Eq.(\ref{eqn_kie_piksystem}) with suitable normalization.}
\label{fig_evln_pik}
\end{figure}
\begin{figure*}
\centering
\subfloat{
   \includegraphics[height=6.0cm,width=7.5cm]{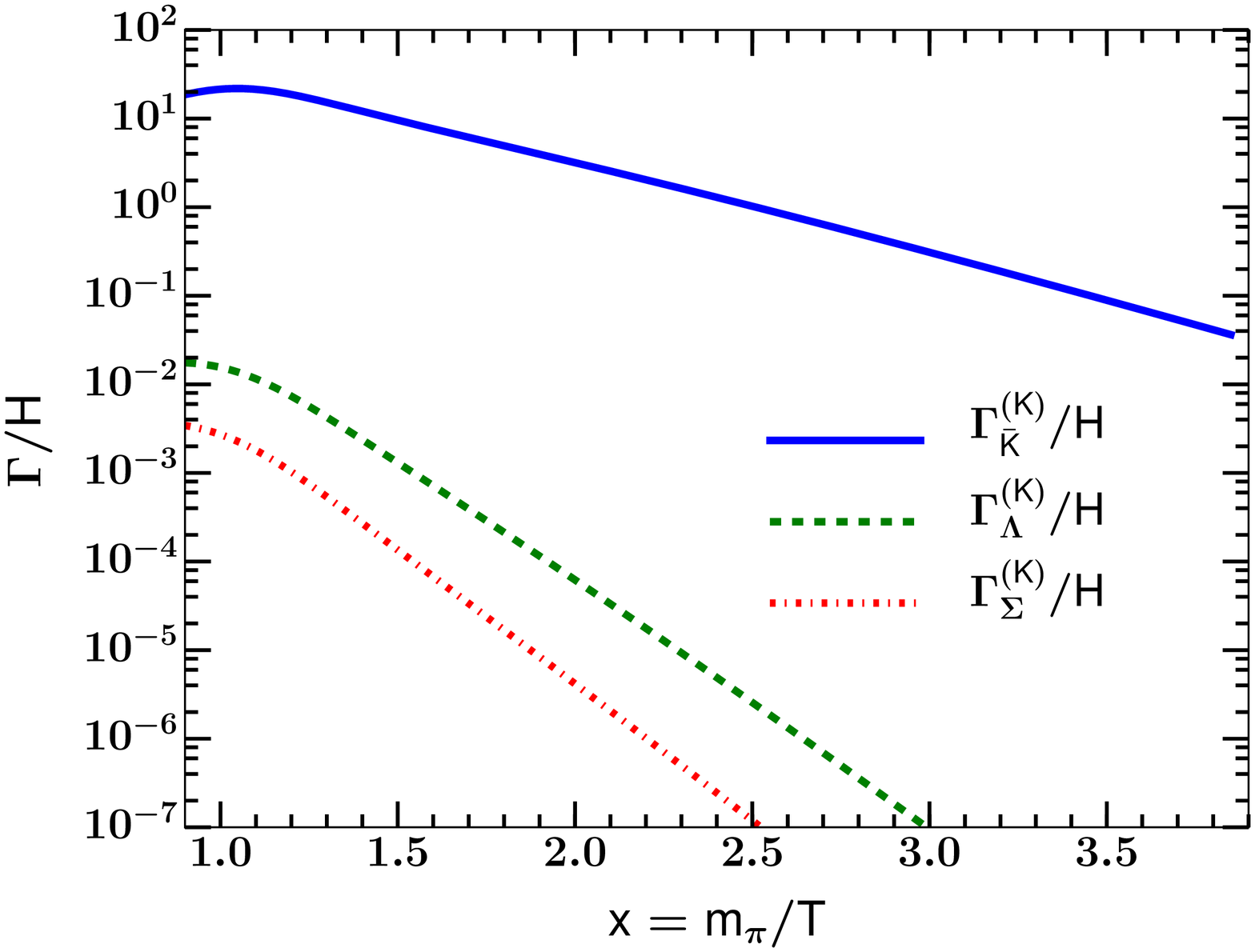}
 }
\subfloat{
   \includegraphics[height=6.0cm,width=7.5cm]{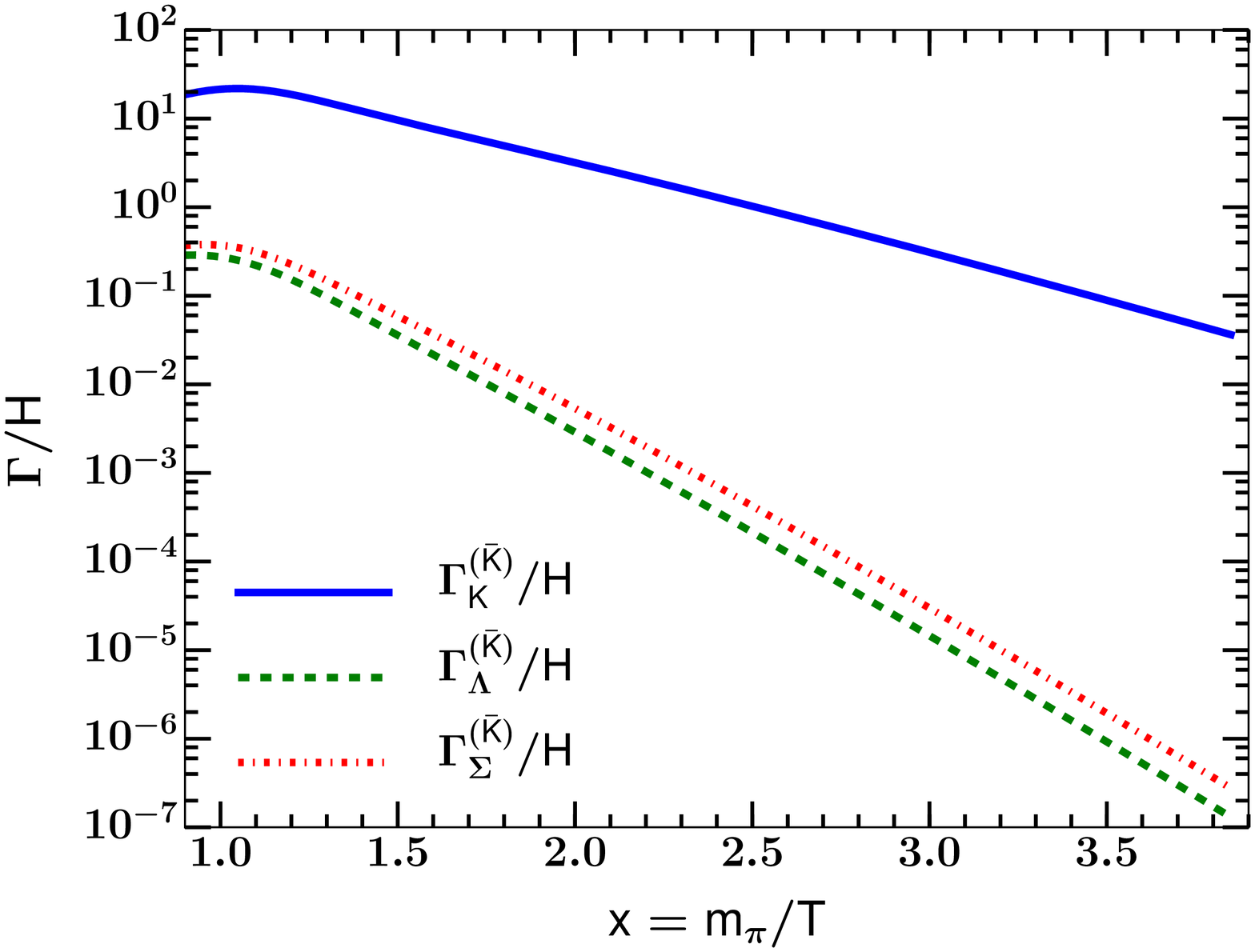}
 }\\
\subfloat{
   \includegraphics[height=6.0cm,width=7.5cm]{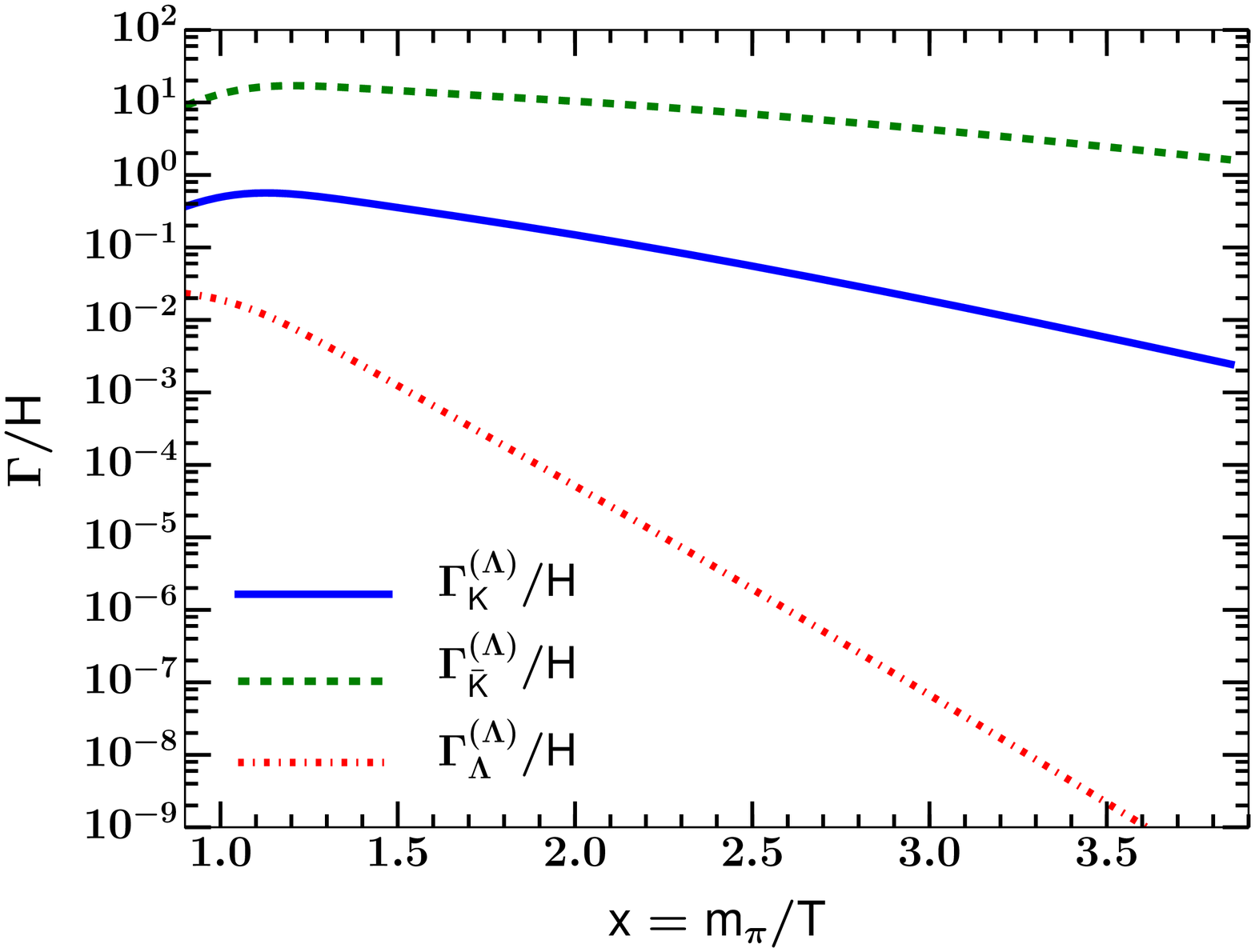}
 }
\subfloat{
   \includegraphics[height=6.0cm,width=7.5cm]{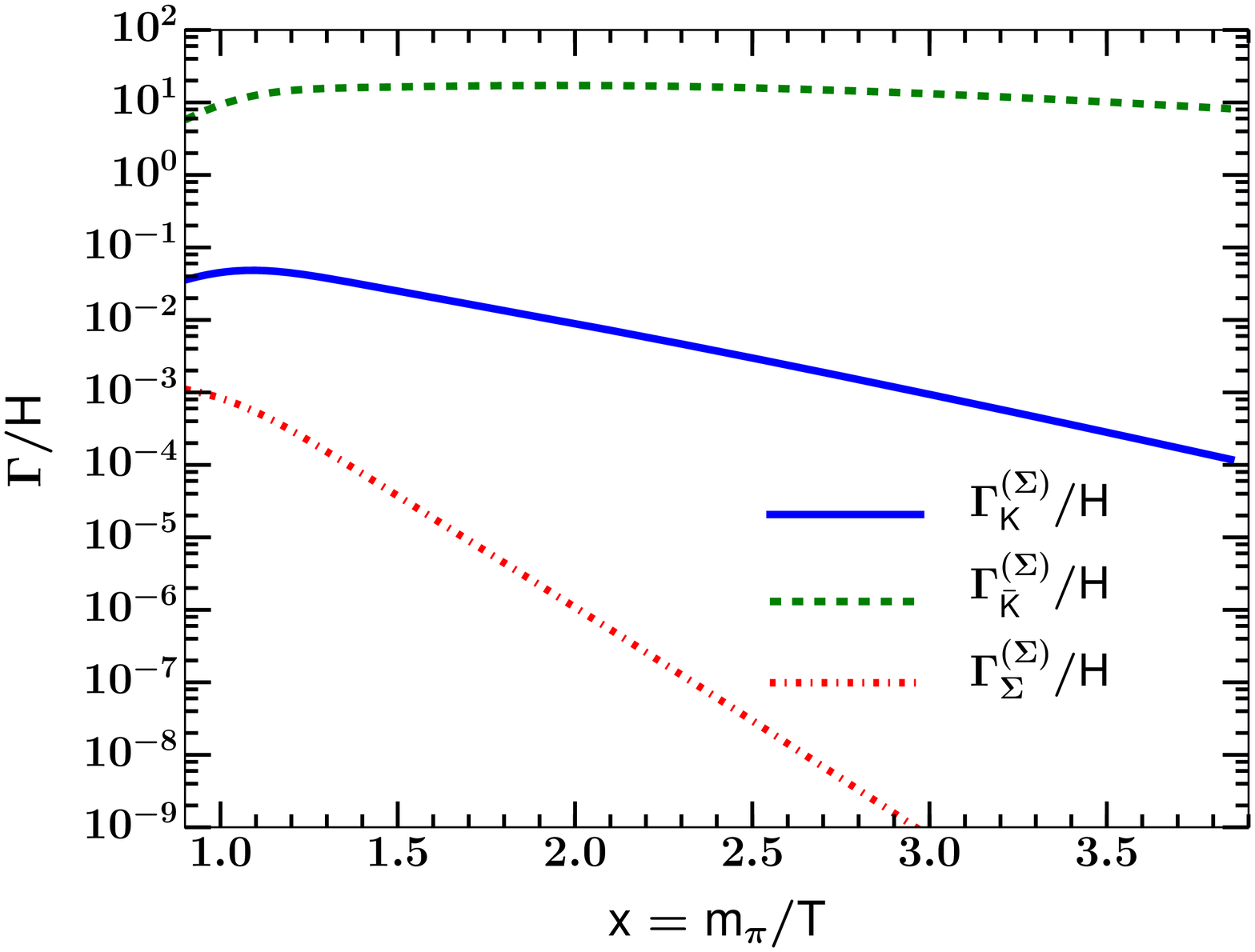}
 } 
\caption{Different rates appearing in Eqs.(\ref{eqn_evln_all_1}-\ref{eqn_evln_all_4}) for BJ case.}
\label{fig_ratio_all_BJ}
\end{figure*}
\begin{figure*}
\centering
\subfloat{
   \includegraphics[height=6.0cm,width=7.5cm]{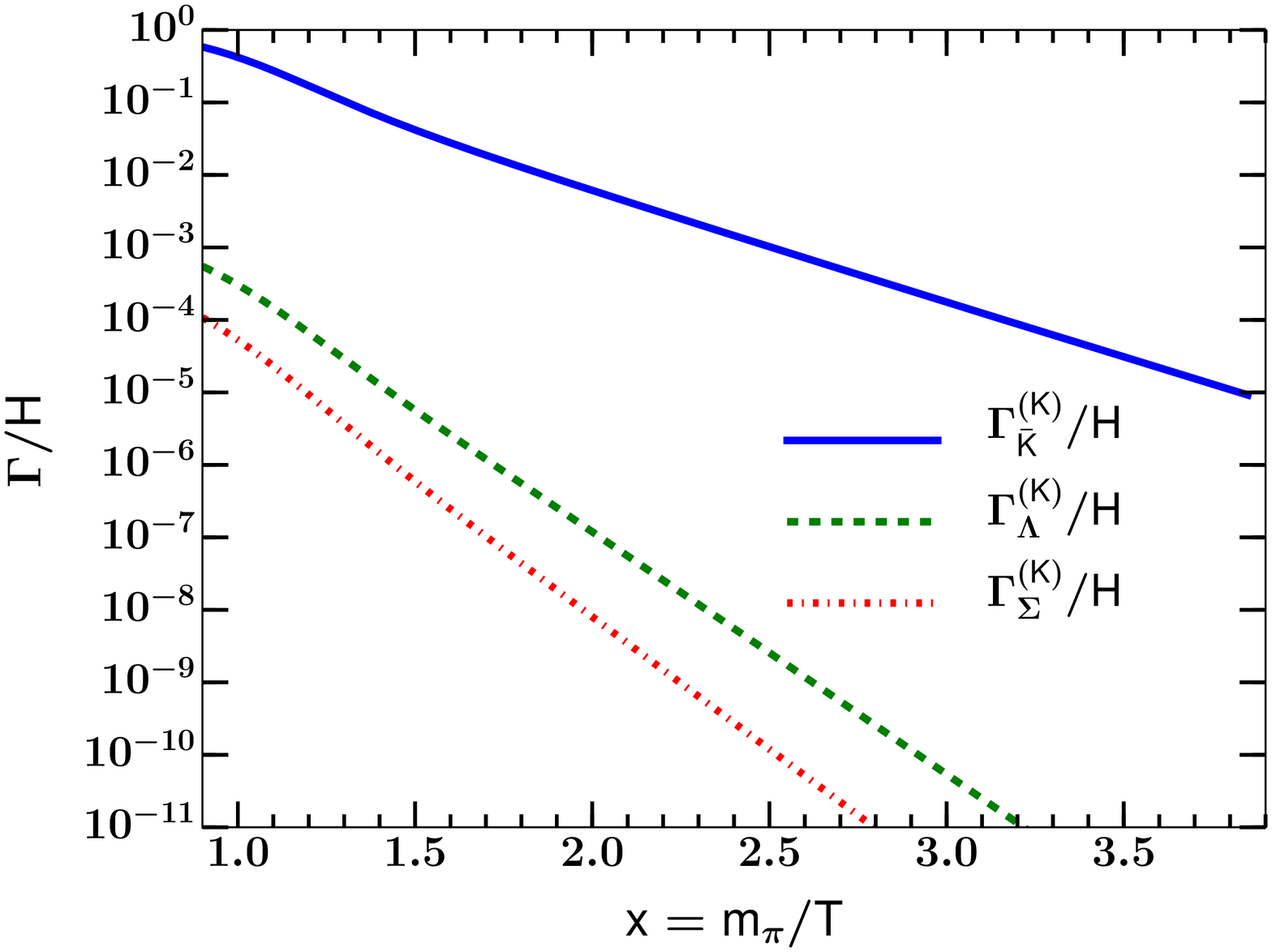}
 }
\subfloat{
   \includegraphics[height=6.0cm,width=7.5cm]{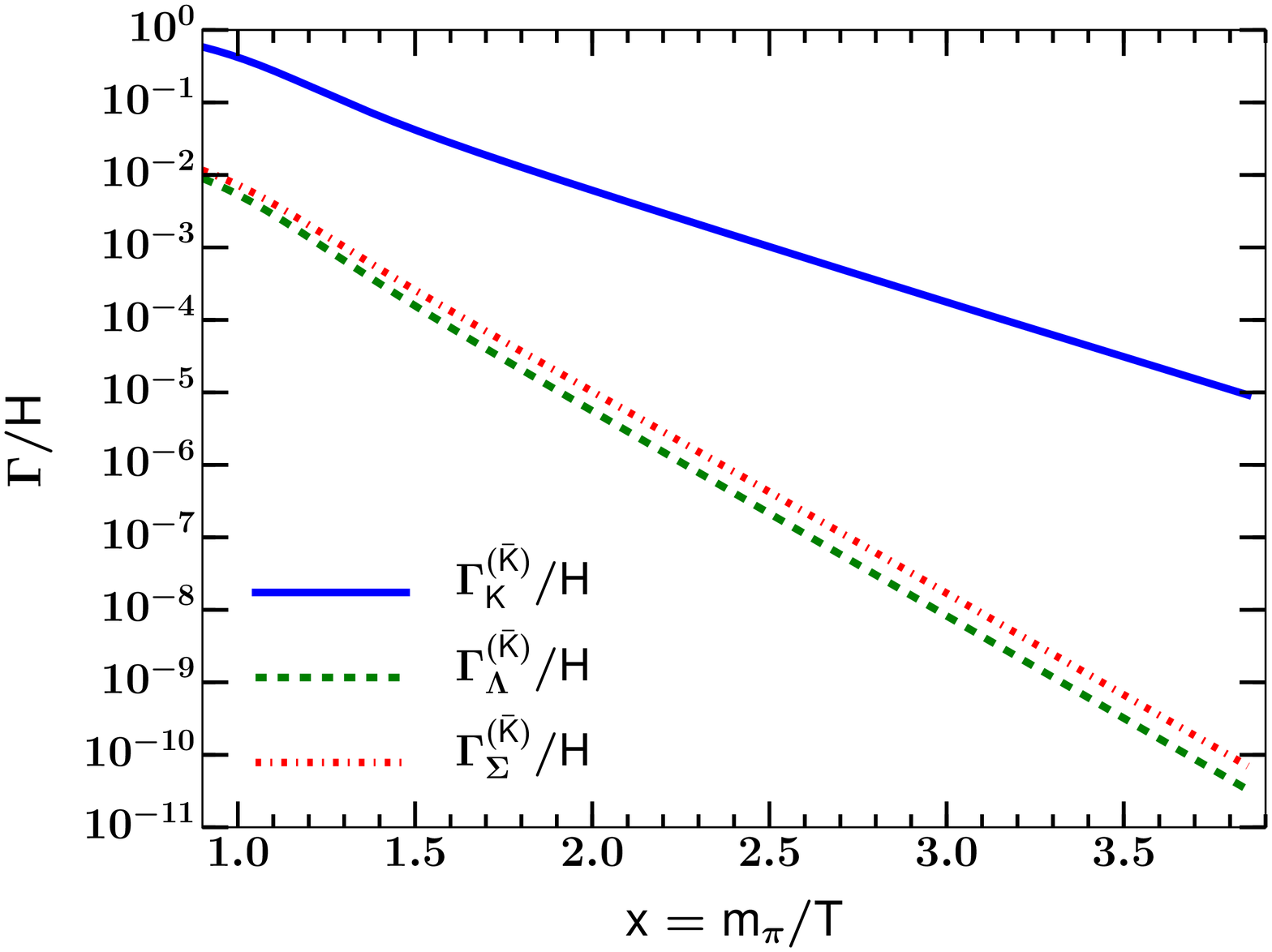}
 }\\
\subfloat{
   \includegraphics[height=6.0cm,width=7.5cm]{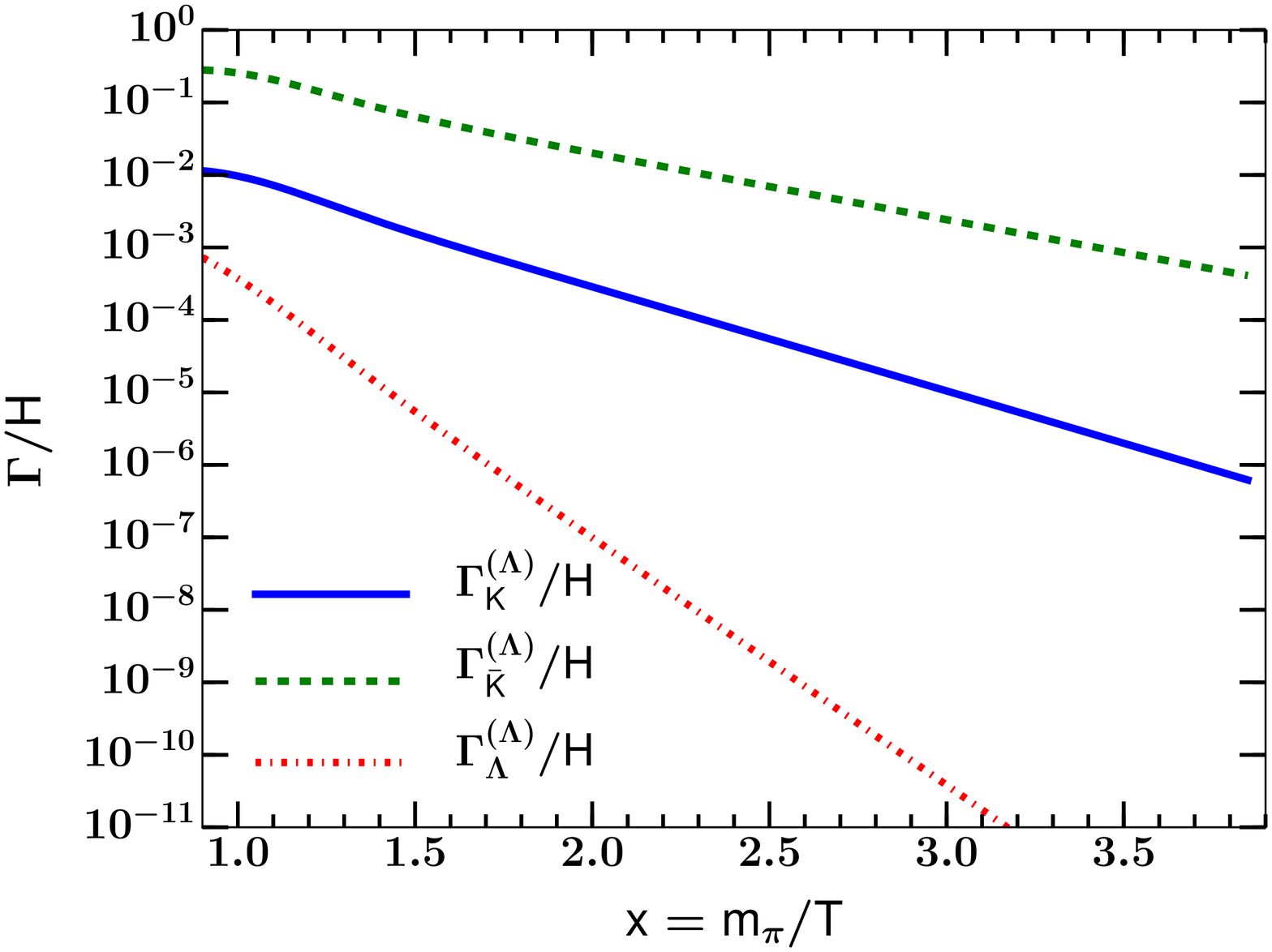}
 }
\subfloat{
   \includegraphics[height=6.0cm,width=7.5cm]{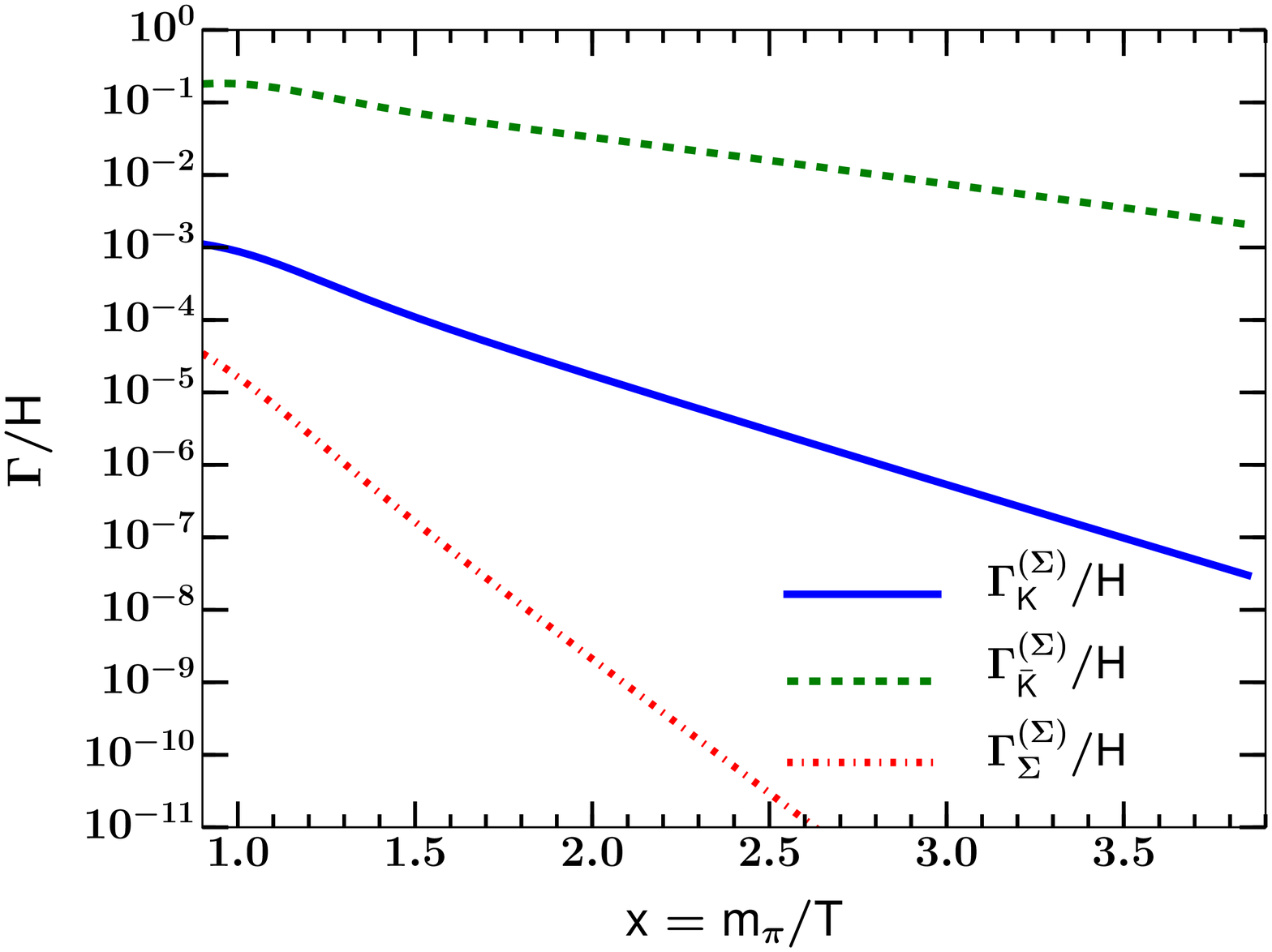}
 } 
\caption{Different rates appearing in Eqs.(\ref{eqn_evln_all_1}-\ref{eqn_evln_all_4}) for HB case.}
\label{fig_ratio_all_HB}
\end{figure*}
\begin{figure*}
\centering
\subfloat{
   \includegraphics[height=6.0cm,width=7.5cm]{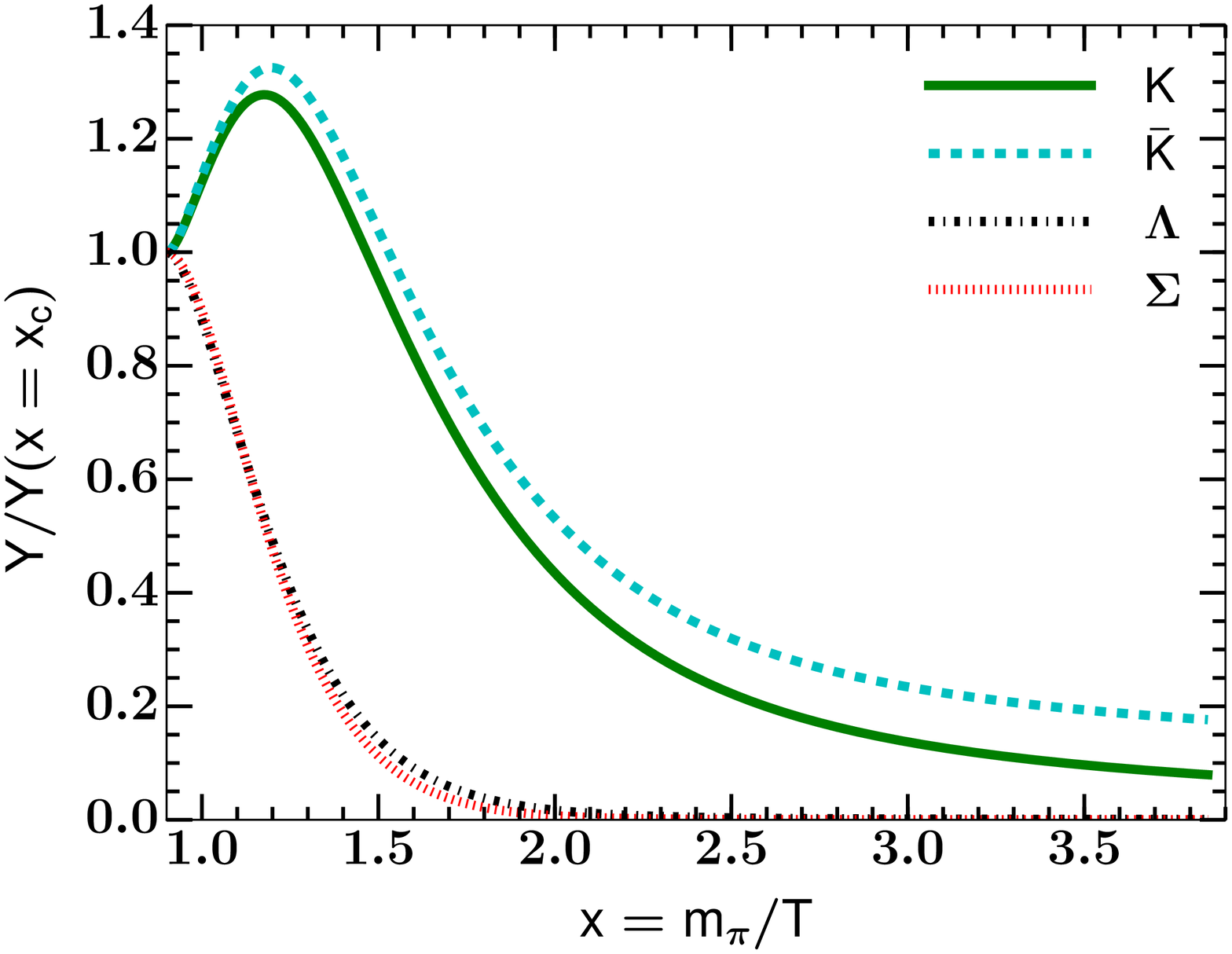}
 }
\subfloat{
   \includegraphics[height=6.0cm,width=7.5cm]{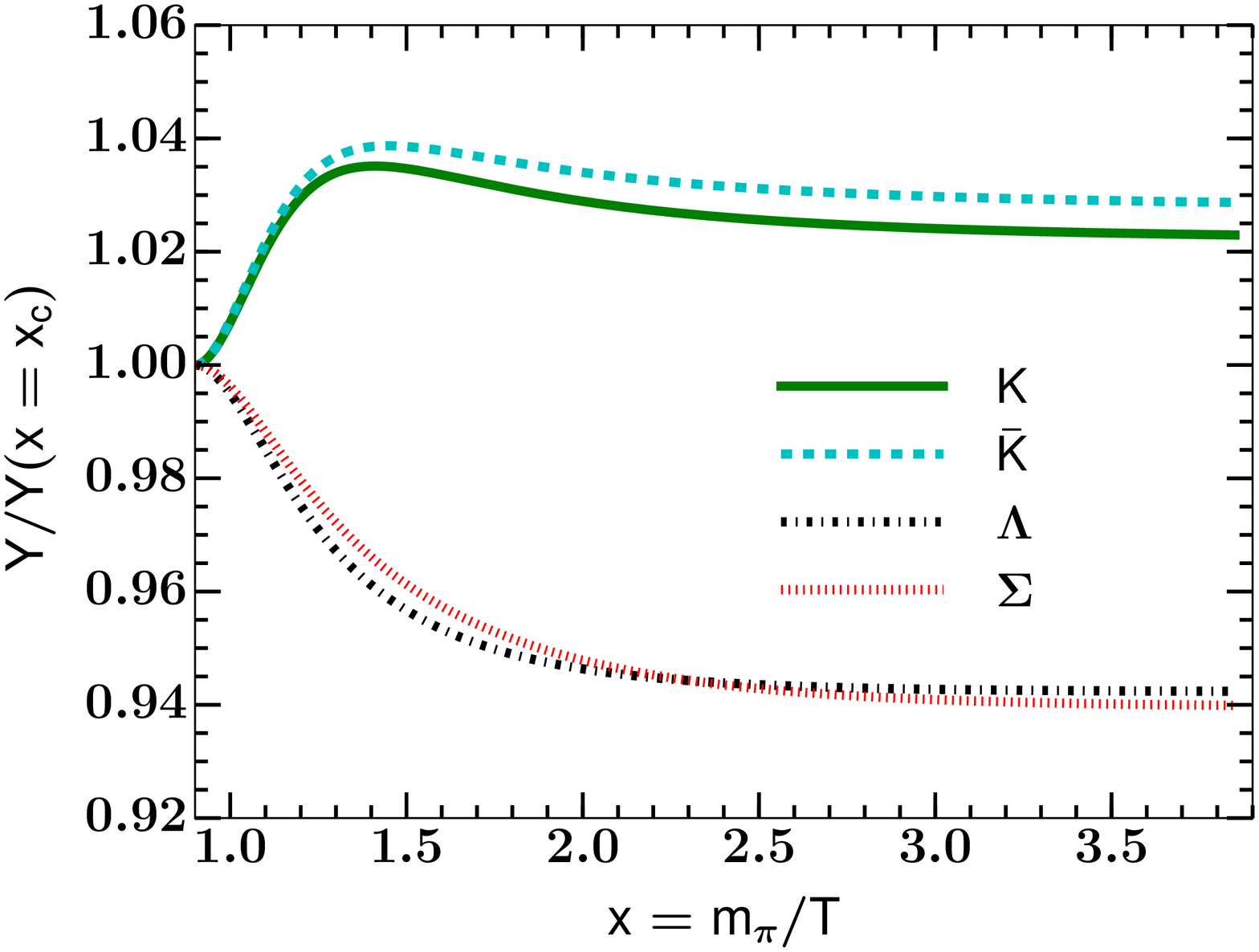}
 } 
\caption{Numerical solution of Eqs.(\ref{eqn_evln_all_1}-\ref{eqn_evln_all_4}) (left panel) BJ case and (right panel) HB case}
\label{fig_plot_evln_all}
\end{figure*}
\begin{figure*}
\centering
\subfloat{
   \includegraphics[height=6.0cm,width=7.5cm]{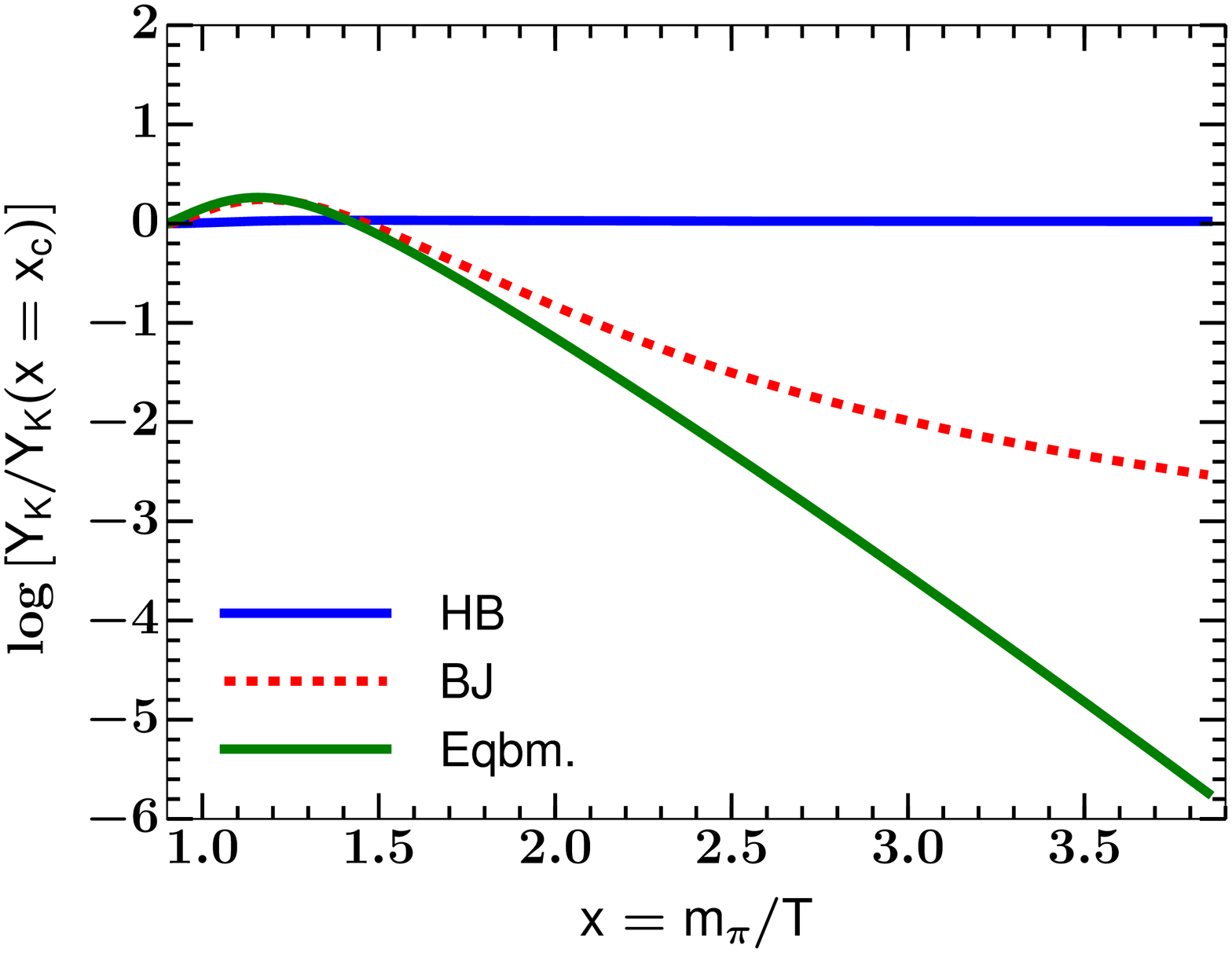}
 }
\subfloat{
   \includegraphics[height=6.0cm,width=7.5cm]{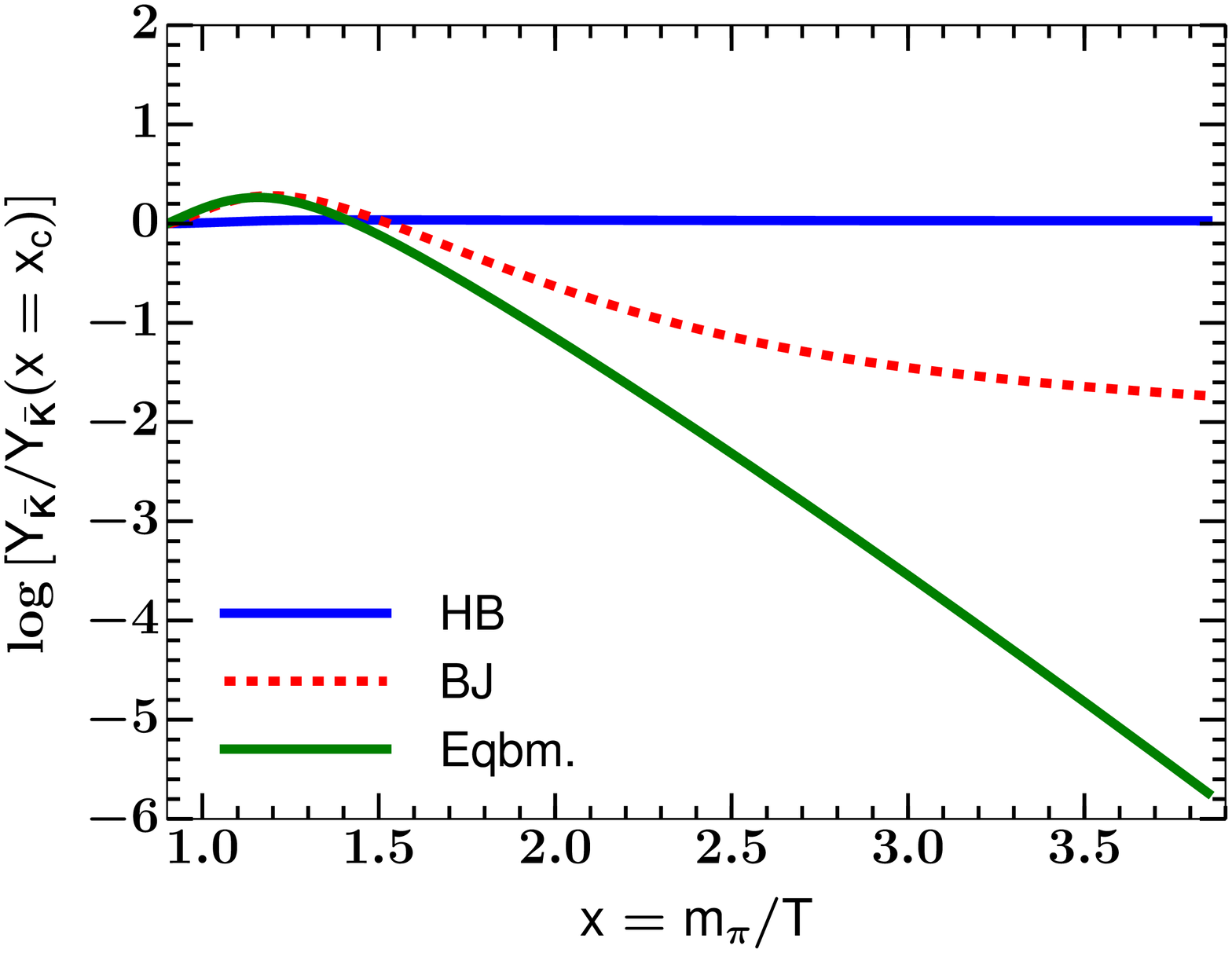}
 }\\
\subfloat{
   \includegraphics[height=6.0cm,width=7.5cm]{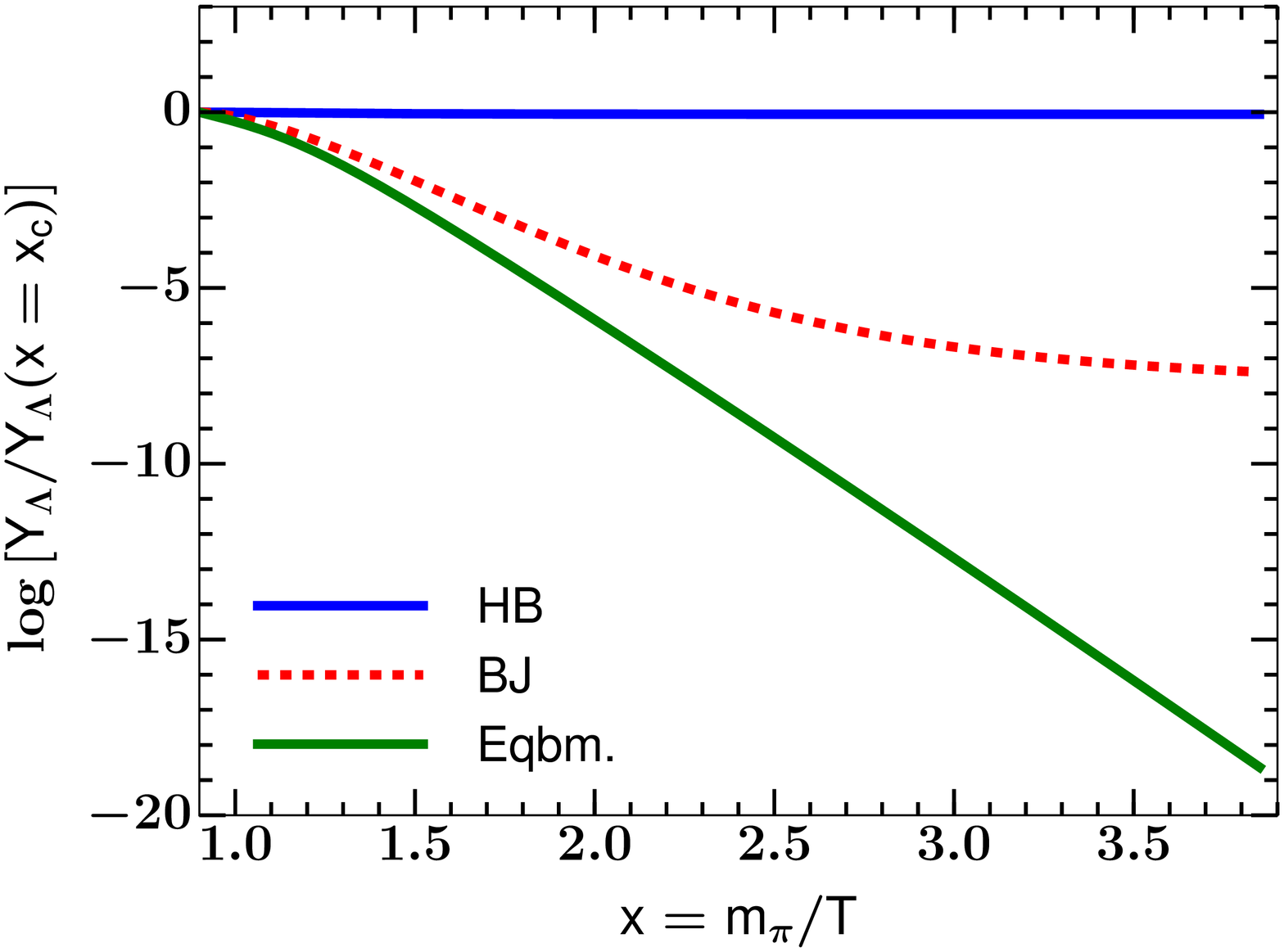}
 }
\subfloat{
   \includegraphics[height=6.0cm,width=7.5cm]{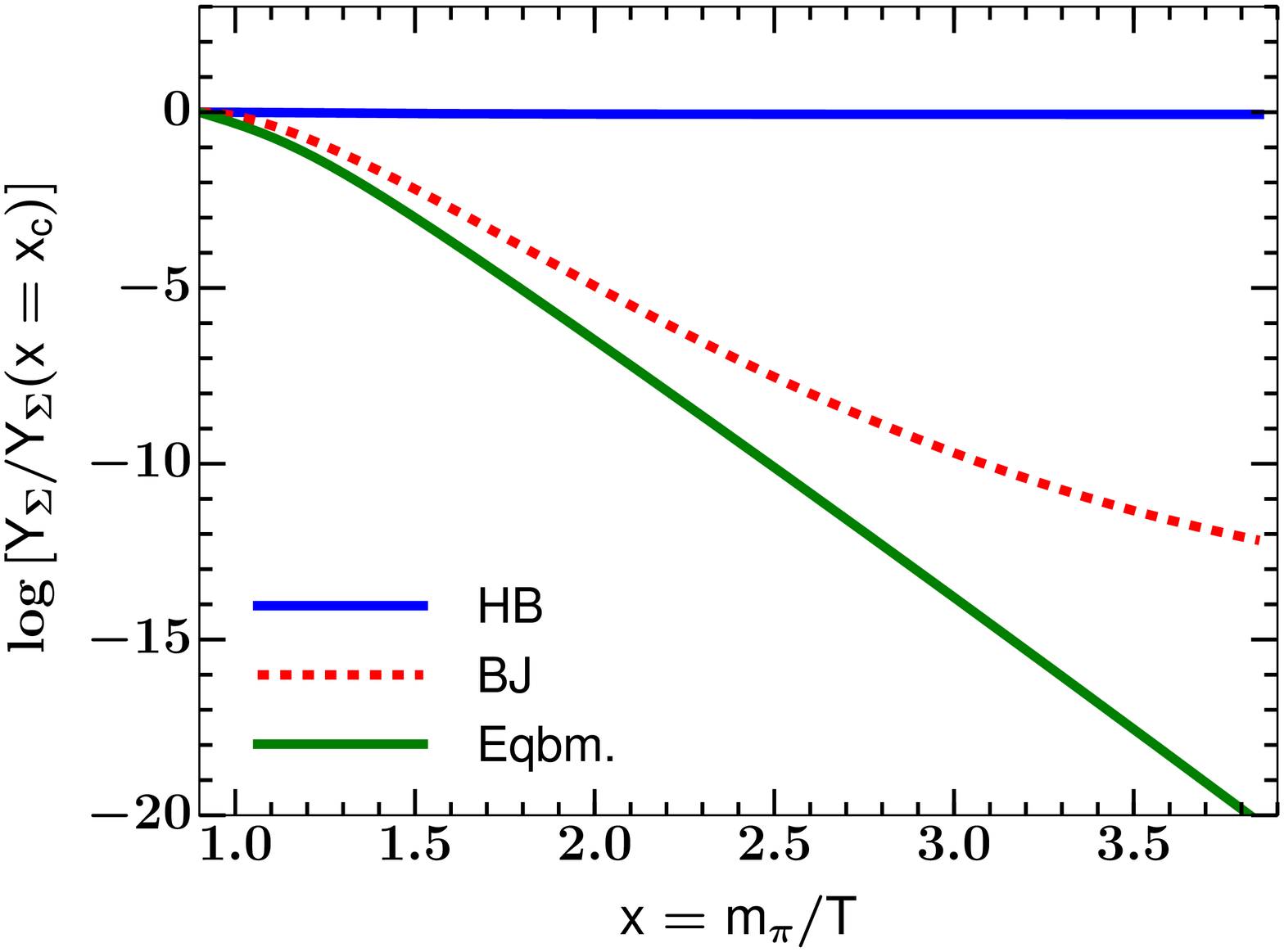}
 } 
\caption{Numerical solution of Eqs.(\ref{eqn_evln_all_1}-\ref{eqn_evln_all_4}) with suitable normalization.}
\label{fig_plot_evln_all_log}
\end{figure*}
Now another question comes vividly, whether all the species freeze out simultaneously? It is well known that the continuous processes of freeze out which starts at $T_{chi}$ and ends at $T_{ch}$ is different for different species as interactions are different. Hence $T_{ch}$ for different species is sequential nature. In this work, this particular aspect is discussed from a microscopic point of view.

Simultaneous or sequential freeze out appears in the discussion when the fluid is multi-component like the hadronic systems produced at RHIC and LHC. When we say, all hadron species freeze out simultaneously at common $T_{ch}$, that means the mean free paths of all species become infinite at that temperature. It may happen in case the fluid expands suddenly, making number density of each species very very low. But in general situation of hadronic matter formed at RHIC and LHC, that probability is less. We have shown the results with extreme conditions of expansion following bjorken and hubble like dynamics and exclude the possibility of simultaneous freeze out condition. 

The aim of the present study is to understand the dynamics of chemical freeze-out of single-strange hadrons ($K,\, \Lambda,\, \Sigma$) from a thermal background provided by non-strange hadrons ($\pi,\, \rho,\,N$). This may be extended to understand the dynamics of other multi strange hadrons. The assumption of non-strange hadrons providing a thermal background can be justified as follows: the non-strange hadrons involve pions which are the lightest hadrons and are produced in large numbers compared to the strange hadrons. Also, $\pi \,N$ cross-sections are usually larger compared to meson-meson interactions and $\rho$-meson is a resonance state in the $\pi\,\pi$ scattering. Hence, the interaction rate among non-strange hadrons, at any time would be larger compared to strange hadrons so that they achieve equilibrium faster compared to strange hadrons. We employ momentum integrated Boltzmann equation or rate equations to study the dynamics. The details of the cross-sections, which go as input in the rate equations, can be found in~\cite{ourprd20}.

After complete chemical freeze-out, the yields of each particle species do not change. However, the elastic scatterings continue, maintaining the kinetic equilibrium till kinetic freeze-out starts. We do not comment anything on kinetic freeze-out as this would require to solve the entire Boltzmann transport equation in an expanding background which we postpone to some future study. In the next section we discuss a very naive $\pi-K$ system and the decoupling of $K$.
\section{\label{sec_pik} $\pi K$ System}
We first consider a system of $\pi$ and $K$-mesons assuming  $n_K = n_{\bar{K}}$ at all times. The only possible inelastic reactions in such a system are $\pi \pi\rightarrow K\bar{K}$ and $K\bar{K}\rightarrow \pi \pi$. Pion is assumed to provide the thermal back ground and $K$ is out of equilibrium. Now to analyse the freeze out, the momentum integrated Boltzmann equation or chemical rate equation for $K$ can be written following Ref.\cite{kolb_book} as
\begin{equation}
 \frac{dn_K}{dt} + \Gamma_{e}n_K = R^o _{\pi \pi\rightarrow K\bar{K}} n_\pi n_\pi -
   R^o _{K\bar{K}\rightarrow \pi \pi }  n_Kn_{\bar{K}}
\end{equation}
where $\Gamma_{e}$ denotes the expansion rate of the system and $R^o_{a b\rightarrow cd}$ denotes the thermal reaction rate as defined in~\cite{kapusta,KbyPi_enhance}
$$R^o_{a b\rightarrow cd} \equiv \langle \sigma_{a b\rightarrow cd}v_{ab}\rangle = \int d^3p_1 \, d^3p_2  \, f^{\text{eq}}_a(p_1)\,f^{\text{eq}}_b(p_2)\, \sigma \,v_{ab} $$
where $\sigma$ denotes the cross-section for the reaction $a b\rightarrow cd$, $f^{\text{eq}}_a,\,f^{\text{eq}}_b$ denote the equilibrium distributions of $a$ and $b$ respectively and $v_{ab}$ denotes the relative velocity (actually M\o{}ller velocity) and is given by
$$v=\frac{\sqrt{(p_a.p_b)^2-m_a^2m_b^2}}{E_aE_b}$$
$p,m$ and $E$ are the momentum, mass and energy of the interacting particles.
If we assume classical Boltzmann statistics for all the particles then the thermal reaction rate can be written as~\cite{KbyPi_enhance}
\begin{widetext}
\begin{equation}
 R^o_{a b\rightarrow cd} (T) = \frac{Tg_ag_b}{32\pi^4 n_a^{\text{eq}}(T) n_b^{\text{eq}} (T)}\int _{s_0}^\infty ds \, [s-(m_a+m_b)^2][s-(m_a-m_b)^2]\, \frac{\sigma(s)}{\sqrt{s}} \, K_1(\sqrt{s}/T)
\end{equation}
\end{widetext}
where $\sqrt{s_0} = \max (m_a+m_b,m_c+m_d)$. In absence of the expansion term, equilibrium condition demands that we must have
$$ R^o _{\pi \pi\rightarrow K\bar{K}} n^\text{eq}_\pi n^\text{eq}_\pi =  R^o _{K\bar{K}\rightarrow \pi \pi }  n^\text{eq}_Kn^\text{eq}_{\bar{K}}$$
and since we assume $\pi$'s to be always in equilibrium, hence the rate equation becomes
\begin{equation}
 \frac{dn_K}{dt} + \Gamma_{e}n_K = R^o _{K\bar{K}\rightarrow \pi \pi }\left[ n^\text{eq}_Kn^\text{eq}_{\bar{K}} - n_Kn_{\bar{K}}\right]
\end{equation}

This equation is valid for both static and expanding system. However, for expanding system the temperature changes along with system dimension as time progresses. To take care of that, both Bjorken and Hubble-like expansion have been considered. For Bjorken flow, the fluid 4-velocity is given as
$$u^\mu = \frac{t}{\tau}\left( 1,0,0,\frac{z}{t}\right)$$
with $\tau = \sqrt{t^2-z^2}$, one gets the expansion rate 
$$\Gamma^B_{e} = \frac{1}{\tau}$$
and for Hubble-like flow, the fluid velocity
$$u^\mu = \frac{t}{\tau}\left( 1,\frac{x}{t},\frac{y}{t},\frac{z}{t}\right)$$
with $\tau =\sqrt{t^2-x^2-y^2-z^2}$. This leads to the expansion rate 
$$\Gamma^H_{e} = \frac{3}{\tau}$$
If the emission is homogeneous and isotropic, then we can set $x=y=z=0$ and get
$$\Gamma_{e} = \left\{ \begin{array}{ll}
           1/t & \text{  Bjorken flow (BJ)}\\
           3/t & \text{  Hubble-like flow (HB)}\\
          \end{array}
 \right.$$
Both dynamics provide the extreme conditions of expansion for a hadronic system. Bjorken describes the slowest whereas Hubble describes the fastest expansion. 

Following~\cite{kolb_book}, we now define $Y_K = n_K/s$, where $s$ is the entropy density, to scale out the effect of expansion and $x=m_\pi/T$ which is a dimensionless quantity. If total entropy is conserved, then we have
$$\frac{dn_K}{dt} + \Gamma_{e}n_K = s\, \dot{Y}_K$$
where, $\Gamma_e=-\frac{1}{s}\frac{ds}{dt}$. Similarly,
$$\dot{Y}_K = \frac{dx}{dt}\frac{dY_K}{dx} = -\frac{m_\pi}{T^2}\left( \frac{dT}{dt}\right)\frac{dY_K}{dx} = -\frac{x^2}{m_\pi}\left( \frac{dT}{dt}\right)\frac{dY_K}{dx}$$
Hence the rate equation becomes
\begin{equation}
 \frac{dY_K}{dx} = -\frac{m_\pi}{s\,x^2}\left( \frac{dT}{dt}\right)^{-1}R^o _{K\bar{K}\rightarrow \pi \pi }\left[ n^\text{eq}_Kn^\text{eq}_{\bar{K}} - n_Kn_{\bar{K}}\right]
\end{equation}
can also be simplified to
\begin{equation}
\label{eqn_kie_piksystem}
 \frac{dY_K}{dx} = -\frac{m_\pi\,s}{x^2}\left( \frac{dT}{dt}\right)^{-1}R^o _{K\bar{K}\rightarrow \pi \pi }\left[ Y^\text{eq}_KY^\text{eq}_{\bar{K}} - Y_KY_{\bar{K}}\right]
\end{equation}
Now to solve the above equation, we need $\frac{dT}{dt}$. Since we have assumed total entropy to be conserved, $\frac{dT}{dt}$ can be obtained by solving the following equation
\begin{equation}
\label{eqn_entropy_continuity}
 \frac{ds}{dt} + \Gamma_{e}s = 0
\end{equation}
Now $s$ is a function of $T$ only. We use the parametrization obtained from Lattice QCD calculation to obtain $s$ as a function of $T$ and consequently obtain $\frac{dT}{dt}$. 
The solution for $T$ and $\frac{dT}{dt}$, after solving Eq.(\ref{eqn_entropy_continuity}), are shown in Fig.(\ref{fig_expansion_model}) as a function of time and temperature respectively.

Having obtained a model for $\frac{dT}{dt}$, we can now solve Eq.(\ref{eqn_kie_piksystem}). But before coming to the result, we write Eq.(\ref{eqn_kie_piksystem}) in a compact form as follows~\cite{kolb_book}:
\begin{align}
 \frac{x}{Y^\text{eq}_K}\frac{dY_K}{dx} &= \frac{n^\text{eq}_{\bar{K}}\,R^o _{K\bar{K}\rightarrow \pi \pi } }{\left( \dfrac{1}{T}\dfrac{dT}{dt} \right)}\left[ \frac{Y_KY_{\bar{K}}}{Y^\text{eq}_KY^\text{eq}_{\bar{K}}}-1\right] \nonumber \\
 &= -\frac{\Gamma }{H}\left[ \frac{Y_KY_{\bar{K}}}{Y^\text{eq}_KY^\text{eq}_{\bar{K}}}-1\right]
 \label{rateeq_m}
\end{align}
where $\Gamma = n^\text{eq}_{\bar{K}}\,R^o _{K\bar{K}\rightarrow \pi \pi }$ denotes the scattering rate, basically here, the rate at which kaons annihilate to pions and $H = \left|\dfrac{1}{T}\dfrac{dT}{dt}\right|$ denotes the rate at which system cools. The rate equation Eq.\ref{rateeq_m} suggests that the scattering rate $\Gamma$ drives the system towards equilibrium whereas $H$ opposes the approach towards equilibrium. The ratio $\Gamma/H$ for $\pi-K$ system is shown in Fig.(\ref{fig_ratio_pik}).

The numerical solution of Eq.(\ref{eqn_kie_piksystem}) is shown in Fig.(\ref{fig_evln_pik}) for both BJ and HB cases. The evolution starts at $T\approx T_c = 155 \text{ MeV}$ or $x_c = m_\pi/T_c$ and the evolution is carried out for a longer time till system has $x=3.85$ (equivalently $T\approx 36\text{ MeV}$). The initial number density of kaon is chosen to be the equilibium value at $T_c$ \emph{i.e.} $n_0 \equiv n^{\text{eq}}(T_c)$, the reason for this choice is to find that if the system formed in equilibrium at $T_c$, then at what time does it fall out-of-equilibium. 

As soon as the evolution starts, the expansion drives the system out-of-equilibium. For the BJ case, however, the ratio $\Gamma/H $ is greater than 1 in the beginning so that $\Gamma$ manages to keep the system close to equilibium. After some time, as $\Gamma$ decreases, $Y^{\text{eq}}_K$ also decreases exponentially, then the system fails to follow the equilibium curve and goes away from equilibium.  However, it can be seen from Fig.\ref{fig_evln_pik} that for the BJ case, the net yield changes slowly and take a longer time for the freeze out process. This is because of the slow expansion in case of Bjorken-dynamics. 

On the other hand, for the HB case, the cooling rate is much faster. The scattering rate, $\Gamma$ becomes lower much earlier than the BJ case compared to expansion rate. Further decrease in scattering rate and with increase in system dimension kaon leads to chemical freeze out where yield gets frozen. Freeze out temperature is close to $T_c$.

\section{\label{sec_all} $\pi\,K\,\rho\,N\,\Lambda,\Sigma$ system}
Now we consider a system consisting of $\pi,\,K,\, \bar{K},\,\rho,\,N,\,\Lambda,\Sigma$. We assume that the non-strange particles $(\pi,\,\rho,\,N)$ provide a thermal background till the time of the freeze-out of the strange hadrons $(K,\, \bar{K},\,\Lambda,\Sigma)$. Considering all possible inelastic reactions and following the notations of the previous section, the evolution equations for $K, \bar{K}, \Lambda$ and $\Sigma$ can be written as 
\begin{widetext}
\begin{align}
 \frac{x}{Y^\text{eq}_K}\frac{dY_K}{dx} &= -\frac{\Gamma^{(K)}_{\bar{K}}}{H}\left[ \frac{Y_KY_{\bar{K}}}{Y^\text{eq}_KY^\text{eq}_{\bar{K}}}-1\right] -\frac{\Gamma^{(K)}_{\Lambda} }{H}\left[ \frac{Y_KY_{\Lambda}}{Y^\text{eq}_KY^\text{eq}_{\Lambda}}-1\right] -\frac{\Gamma^{(K)}_{\Sigma}}{H}\left[ \frac{Y_KY_{\Sigma}}{Y^\text{eq}_KY^\text{eq}_{\Sigma}}-1\right] \label{eqn_evln_all_1}\\
 \frac{x}{Y^\text{eq}_{\bar{K}}}\frac{dY_{\bar{K}}}{dx} &= -\frac{\Gamma^{(\bar{K})}_{K}}{H}\left[ \frac{Y_{\bar{K}}Y_K}{Y^\text{eq}_{\bar{K}}Y^\text{eq}_K}-1\right] -\frac{\Gamma^{(\bar{K})}_{\Lambda} }{H}\left[ \frac{Y_{\bar{K}}}{Y^\text{eq}_{\bar{K}}}-\frac{Y_{\Lambda}}{Y^\text{eq}_{\Lambda}}\right] -\frac{\Gamma^{(\bar{K})}_{\Sigma}}{H}\left[ \frac{Y_{\bar{K}}}{Y^\text{eq}_{\bar{K}}}-\frac{Y_{\Sigma}}{Y^\text{eq}_{\Sigma}}\right] \label{eqn_evln_all_2}\\
 \frac{x}{Y^\text{eq}_\Lambda}\frac{dY_\Lambda}{dx} &= -\frac{\Gamma^{(\Lambda)}_{K}}{H}\left[ \frac{Y_\Lambda Y_{K}}{Y^\text{eq}_\Lambda Y^\text{eq}_{K}}-1\right] - \frac{\Gamma^{(\Lambda)}_{\bar{K}} }{H}\left[ \frac{Y_{\Lambda}}{Y^\text{eq}_{\Lambda}}-\frac{Y_{\bar{K}}}{Y^\text{eq}_{\bar{K}}}\right] -\frac{\Gamma^{(\Lambda)}_{\Lambda}}{H}\left[ \frac{\left(Y_\Lambda\right)^2}{\left( Y^\text{eq}_\Lambda \right)^2}-1\right] \label{eqn_evln_all_3}\\
 \frac{x}{Y^\text{eq}_\Sigma}\frac{dY_\Sigma}{dx} &= -\frac{\Gamma^{(\Sigma)}_{K}}{H}\left[ \frac{Y_\Sigma Y_{K}}{Y^\text{eq}_\Sigma Y^\text{eq}_{K}}-1\right] - \frac{\Gamma^{(\Sigma)}_{\bar{K}}}{H}\left[ \frac{Y_\Sigma }{Y^\text{eq}_\Sigma}-\frac{Y_{\bar{K}}}{ Y^\text{eq}_{\bar{K}}}\right] -\frac{\Gamma^{(\Sigma)}_{\Sigma} }{H}\left[ \frac{\left(Y_\Sigma\right)^2}{\left(Y^\text{eq}_\Sigma\right)^2 }-1\right] \label{eqn_evln_all_4}
\end{align}
\end{widetext}
the scattering rates of various channels contributing to the net productions are defined as follows. 
\begin{align*}
 \Gamma^{(K)}_{\bar{K}} &= n^\text{eq}_{\bar{K}}\left[ R^o_{K\bar{K}\rightarrow \pi \pi} + R^o_{K\bar{K}\rightarrow \pi \rho} + R^o_{K\bar{K}\rightarrow \rho \rho} + R^o_{K\bar{K}\rightarrow p \bar{p}}\right]\\
 \Gamma^{(K)}_{\Lambda} &= n^\text{eq}_{\Lambda}\left[ R^o_{K\Lambda \rightarrow \pi N} + R^o_{K\Lambda \rightarrow \rho N}\right]\\
 \Gamma^{(K)}_{\Sigma} &= n^\text{eq}_{\Sigma}R^o_{K\Sigma \rightarrow \pi N}\\
 \Gamma^{(\bar{K})}_{K} &= n^\text{eq}_{K}\left[ R^o_{K\bar{K}\rightarrow \pi \pi} + R^o_{K\bar{K}\rightarrow \pi \rho} + R^o_{K\bar{K}\rightarrow \rho \rho} + R^o_{K\bar{K}\rightarrow p \bar{p}}\right]\\
 \Gamma^{(\bar{K})}_{\Lambda} &= n^\text{eq}_N R^o_{\bar{K}N \rightarrow \Lambda\pi } \\
 \Gamma^{(\bar{K})}_{\Sigma} &= n^\text{eq}_N R^o_{\bar{K}N \rightarrow \Sigma\pi }\\
 \Gamma^{(\Lambda)}_{K} &= n^\text{eq}_{K}\left[ R^o_{\Lambda K \rightarrow \pi N} + R^o_{\Lambda K\rightarrow \rho N}\right]\\
 \Gamma^{(\Lambda)}_{\bar{K}} &= n^\text{eq}_\pi R^o_{ \Lambda\pi \rightarrow \bar{K}N }\\
 \Gamma^{(\Lambda)}_{\Lambda} &= n^\text{eq}_{\Lambda} R^o_{\Lambda\Lambda \rightarrow p \bar{p}}\\
 \Gamma^{(\Sigma)}_{K} &= n^\text{eq}_{K}R^o_{\Sigma K \rightarrow \pi N}\\
 \Gamma^{(\Sigma)}_{\bar{K}} &= n^\text{eq}_\pi R^o_{ \Sigma\pi \rightarrow \bar{K}N }\\
 \Gamma^{(\Sigma)}_{\Sigma} &= n^\text{eq}_{\Sigma} R^o_{\Sigma\bar{\Sigma} \rightarrow p \bar{p}}
\end{align*}
The above rates are plotted in Figs.\ref{fig_ratio_all_BJ} \&\ref{fig_ratio_all_HB}.  
The cross sections for various hadronic processes producing hyperons and strange mesons are already mentioned in \cite{ourprd20}. The cross-sections of all inverse reactions are obtained using principle of detailed balance as follows
\begin{equation}
  \sigma_{f\rightarrow i}=\frac{{p_i}^2}{{p_f}^2}\frac{g_i}{g_f}\sigma_{i\rightarrow f}
 \end{equation}
where $p_i,\, p_f$ are the center of mass momenta and $g_i,\, g_f$ are the total degeneracies of the initial and final channels. 

It can be observed that the ratio $\Gamma/H$ is different for different channels. As a result, different channels will freeze-out at different temperatures.

Now we solve the coupled differential equations Eqs.(\ref{eqn_evln_all_1}-\ref{eqn_evln_all_4}) numerically. The evolution starts at $T\approx T_c = 155 \text{ MeV}$ or $x_c = m_\pi/T_c$. The initial number densities are chosen as the equilibium values at $T_c$ \emph{i.e.} $n^{(0)}_K \equiv n^{\text{eq}}_K(T_c),\,n^{(0)}_{\bar{K}} \equiv n^{\text{eq}}_{\bar{K}}(T_c),\,n^{(0)}_\Lambda \equiv n^{\text{eq}}_\Lambda(T_c),\,n^{(0)}_\Sigma \equiv n^{\text{eq}}_\Sigma(T_c)$. The numerical solution is shown in the left and right panel of Fig.(\ref{fig_plot_evln_all}) for the BJ case and HB case respectively and also in Fig.(\ref{fig_plot_evln_all_log}). For the BJ case, $K$ and $\bar{K}$ decouple at later time (or lower temperature) whereas $\Lambda$ and $\Sigma$ decouple earlier (or higher temperature). For the HB case, since the expansion rate is larger, all channels freeze-out as soon as the evolution starts so that the yields of all the particles are fixed at $T\approx T_c$ (the yield however changes by about 2\% for $K, \bar{K}$ and about 6\% for $\Lambda, \Sigma$), hence the chemical freeze-out temperature for all is $T_{\text{ch}}\approx T_c$. But if one observes the value of yield of different species are fixed at different temperatures although close to $T_c$. Again it is prominent if one starts with initial densities little away from equilibrium values.
\section{\label{sec_conclude} Summary and Conclusions}
Hydrodynamics along with transport approach provide a good description of matter created at top RHIC and LHC energies. To infer the accurate properties, it is important to know the  chemical($T_{ch}$) and kinetic ($T_k$) freeze out temperatures  when various particle species decouple chemically and kinetically from the system. 

In this article we have done a microscopic analysis for the chemical freeze-out of single strange hadrons $K, \Lambda$ and 
$\Sigma$. Momentum integrated Boltzmann equation or rate equation has been employed to study the nature of chemical freeze out of these hadrons considering both slowest and fastest expansion dynamics. First the freeze out of kaons in a $\pi-K$ expanding system is studied by comparing the scattering and expansion rate. It has been observed that kaon freeze out takes a longer time  when the system obeys Bjorken dynamics compared to Hubble-like expnasion. In case of Hubble expansion, kaon yield is freezed and the particles decouple chemically near about $T_c$ when the system starts evolving with initial equilibrium kaon density at $T_c$. The same calculation has been extended to study the $K, \Lambda$ and $\Sigma$ freeze out in a system with $\pi,\rho,K, \bar{K}, N,\Lambda, \Sigma$ as constituents. Similar observation of late freeze out for these strange hadrons is obtained in case of Bjorken expansion. However, in case of Hubble like expnasion the strange hadrons freeze out near $T_c$ when hadronic system starts with initial equilibrium density from $T_c$ and evolves further.

It appears that all the strange hadrons do freeze out near $T_c$ which is like simultaneous freeze out of different species.
For faster expansion (here HB case), which is more realistic for the late stages of the fireball formed at RHIC and LHC, that the relaxation rates are much smaller compared to expansion rate right from the beginning which leads to the conclusion that the chemical freeze-out temperature is $T_c$. This is true for Kaon, Lambda and Sigma. 

However, the ratios of scattering rate of strange species to expansion rate of the system suggests that the chemical freeze out is not simultaneous rather sequential. If the strange hadrons are evolved with initial densities away from equilibrium, then it distinguishes different freeze out temperatures for different species. Any realistic calculation would have the expnasion in between these two dynamics. The freeze out process for any particular species is a continuous one, which starts at a temperature $T_{chi}$ and ends at $T_{ch}$. 

Whatever we have discussed is for chemical freeze out as study of kinetic freeze out is more complex using Boltzmann equation where the momentum distribution has to be evolved. 


Simultaneous or sequential freeze out are discussed while dealing with multi component fluid like the hadronic systems produced at top RHIC and LHC energies. If all hadron species freeze out simultaneously at a common temperature then we can have simultaneous freeze out or common universal freeze out temperature, as suggested by Heinz \cite{heinz2006} where mean free paths of all species become infinite. It may only happen in case the fluid expands suddenly, making number density of each species very very low. 
\twocolumngrid

\end{document}